\documentclass[aps,prl,twocolumn,groupedaddress]{revtex4}

\usepackage{graphicx}
\begin{document}

% Use the \preprint command to place your local institutional report
% number in the upper righthand corner of the title page in preprint mode.
% Multiple \preprint commands are allowed.
% Use the 'preprintnumbers' class option to override journal defaults
% to display numbers if necessary
%\preprint{}
%Title of paper
\title{Signatures of Fermion Pairing with Unconventional Symmetry around BCS-BEC Crossover in a Quasi-2D Lattice}

% repeat the \author .. \affiliation  etc. as needed
% \email, \thanks, \homepage, \altaffiliation all apply to the current
% author. Explanatory text should go in the []'s, actual e-mail
% address or url should go in the {}'s for \email and \homepage.
% Please use the appropriate macro foreach each type of information

% \affiliation command applies to all authors since the last
% \affiliation command. The \affiliation command should follow the
% other information
% \affiliation can be followed by \email, \homepage, \thanks as well.
\author{Du\v san Vol\v cko}
\author{Khandker F. Quader}
%\email[]{Your e-mail address}
%\homepage[]{Your web page}
%\thanks{}
%\altaffiliation{}
\affiliation{Department of Physics, Kent State University, Kent, OH
44242}

%Collaboration name if desired (requires use of superscriptaddress
%option in \documentclass). \noaffiliation is required (may also be
%used with the \author command).
%\collaboration can be followed by \email, \homepage, \thanks as well.
%\collaboration{}
%\noaffiliation

\date{\today}

\begin{abstract}
% insert abstract here
We consider fermions on a 2D square lattice with a finite-range 
pairing interaction, and obtain signatures for 
unconventional pair-symmetry states, $d_{x^2-y^2}$ and extended-s ($s^*$), 
in the BCS-BEC crossover region. 
%Starting from s on a 2D lattice using a finite-range 
%pairing interaction derivable from an extended Hubbard model. We obtain signatures of 
%unconventional pair-symmetry states, $d_{x^2-y^2}$ and extended-s ($s^*$), 
%in the BCS-BEC crossover region. 
We find that the fermion momentum distribution function, $v_k^2$, the ratio of the Bogoliubov coefficients,  $v_k/u_k$, and the Fourier transform of $v_k^2$ are strikingly different for $d$- and $s^*$ symmetries in the crossover region. The chemical potential and the gaps functions for both pairing symmetries show several interesting features as a function of interaction. Fermionic atoms in 2D optical lattices may provide a way to test these signatures.
We discuss current generation cold atom experiments that may be utilized. 

\end{abstract}

% insert suggested PACS numbers in braces on next line
% insert suggested keywords - APS authors don't need to do this
%\keywords{}

%\maketitle must follow title, authors, abstract, \pacs, and \keywords
\maketitle

% body of paper here - Use proper section commands
% References should be done using the \cite, \ref, and \label commands
%\section{\label{sec:level1}Introduction}
% Put \label in argument of \section for cross-referencing
%\section{\label{}}

%Among the exciting developments in recent years are discoveries of several classes of 
In recent years there have been fascinating discoveries of several classes of 
systems exhibiting many-body states with paired fermions, that may have unconventional pairing symmetry, different from that in s-wave BCS case. These range from heavy fermions, high-T$_c$  cuprates, iron-pnictides to ultracold fermions. There  are intense theoretical and experimental efforts aimed at deciphering pairing symmetries or mechanisms in these systems.

Ultracold neutral atoms~\cite{bei} present an unprecedented opportunity to study the physics of quantum many-particle systems. Subjected to positive and negative detuning using Feshbach resonance technique, these  provide realizations of weak-coupling BCS to strong coupling BEC crossover behavior, and the unitarity limit, where the scattering length is infinite. 

Cold fermionic atoms in {\it optical lattices}~\cite{greiner02,kohl05} constitute an intriguing set of systems. Tunibility of filling (particles  per site),  hopping kinetic energy or interparticle interaction render optical lattices unprecedented testing grounds for models of correlated electron systems. 
It has been suggested~\cite{jaksch98} that atoms in optical lattices, confined 
to the lowest Bloch band, can be represented by the archetypal condensed matter Hamiltonian, the Hubbard model with 
hopping $t$ between neighboring sites, and on-site interaction $U$. 
Interatomic magnetic or electric dipole interactions, and possibly multi-band couplings, can result in finite-range interactions, that could simulate the extended Hubbard model~\cite{extended}.
%Recent work~\cite{kohl05,duan05,multiband} have pointed out possible
%role of additional Bloch bands and multi-band couplings 
%in optical lattices. In solids this would correspond to having multiple orbitals and
%near-neighbor interactions. wherein it has been suggested~\cite{kotliar88} that t-J model
%can give rise to d-wave pairing.
Duan~\cite{duan05} has shown that on different sides of a 
broad Feshbach resonance, the effective Hamiltonian can be reduced to a t-J model, familiar in studies of correlated electron systems, $J$ being related to spin or magnetic coupling.  Calculations~\cite{micnas,theory,kotliar88} predict that attractive-U Hubbard model gives rise to s-wave superconductivity, while the repulsive-U model results in an antiferromagnetic or a $d$-wave superconducting phase depending on filling.
There has been suggestion~\cite{hofstetter02} that the
underlying physics of the high $T_c$ superconductors may be understood by 
studying optical lattice systems. Also, due to the possibility that the cuprates, possessing short coherence lengths, could fall in the BEC-BCS crossover region, the crossover 
problem~\cite{eagles,leggett} has received considerable theoretical attention.
%~\cite{micnas,VQ,becbcshtccont,melo,becbcshtclat,derhertog99,andrenacci99}.
Several authors employed continuum 
models~\cite{micnas,VQ,becbcshtccont,melo,andrenacci99},
focussing mostly on conventional s-wave pair symmetry. Lattice models with on-site
or nearest-neighbor attractions have also been 
considered~\cite{micnas,VQ,andrenacci99,becbcshtclat,derhertog99}. There is also a large body of theory work~\cite{becbcscold,trivedi} specific to cold fermions.

Observation of Mott-insulator behavior~\cite{mott} and s-wave superfluidity (for attractive interactions)~\cite{ketterle-sf} in 3D optical lattices represent remarkable feats. Current searches for new phases, such as antiferromagnetism (AFM) or unconventional pairing are greatly facilitated by recent  development of analogs to existent powerful experimental tools in condensed matter physics; examples are momentum resolved radio frequency (rf) and rf pairing gap spectroscopies~\cite{rf,rf-pair}, tomographic rf  spectroscopy~\cite{tomog}, out of lattice time-of-flight measurements~\cite{tof}, fluorescence imaging~\cite{fluo}, and momentum-resolved photoemission spectroscopy~\cite{feld}, analogous to angle resolved photoemission spectroscopy (ARPES) in condensed matter~\cite{Shen}.
%thus provide ways to gain useful insight into emergent phenomena in correlated fermi %systems. 

In this paper, we study zero temperature fermion pairing in a 2D square lattice
in the BEC-BCS crossover regime using a finite-range {\it pairing} interaction.
%obtainable from a multi-band {\it extended} Hubbard model. 
As representative cases of unconventional pair symmetry,
we consider two even-parity representations of the
cubic group, namely the $\ell =2$ $d_{x^2-y^2}$-wave, and the
$\ell=0$ extended s-wave ($s^*$). 
%There has been work~\cite{derhertog99,andrenacci99,chen} employing similar pairing %interaction, but their focus is different from ours. 
We present several new results, in particular, specific signatures of states with 
{\it unconventional pairing gap}  symmetries as one goes between weak and strong coupling regimes. We expect our results to be of relevance to fermionic atoms in 2D optical lattices, and to correlated fermion model systems these would simulate. 
%hus to possibly high $T_c$ cuprates.
A key result is
the remarkable behavior of the fermion distribution function, $v_k^2$,
(related to momentum distribution, $n_k$):
For a $d$-wave pairing gap function,  $v_k^2$ changes {\it abruptly} from
exhibiting a peak at the Brillouiun zone (BZ) center (0,0)
to a vanishing central peak accompanied by a
redistribution of the weight around other parts of the
BZ ($(0,\pm \pi)$,$(\pm \pi, 0)$) as the system crosses from
the weak-coupling BCS to the strong-coupling BEC regime. Its Fourier transform 
exhibits a ``checkerboard''  pattern in real space.
By contrast, $v_k^2$ changes smoothly in the $s^*$-wave case.
We find similar signatures in the ratio of Bogoliubov coefficients $v_k/u_k$,
related to the phase of the superfluid wavefunction. 

Our finite-range pairing interaction is obtained from the extended Hubbard model for two equal species population system on a 2D square lattice:
\begin{eqnarray}
H&=&\sum_{<ij>\sigma}(-t c_{i\sigma}^+c_{j\sigma}+\rm{H.c.}) + U\sum_{i}n_{i\sigma}n_{i-\sigma}\nonumber\\
&-&V\sum_{<ij>\sigma\sigma^{\prime}}n_{i\sigma} n_{j\sigma^{\prime}} - \mu_{o}\sum_{i}n_{i},
\end{eqnarray}
where $t$ is the hopping, 
$\mu_{o}$ the unrenormalized chemical potential,
$U$ the on-site repulsion and $V$ the nearest-neighbor attraction.
%In the case of
%cold fermions on a lattice, $V$ would be related to inter-band coupling.
$\sigma$ is the ``spin'' index, that could refer to hyperfine states
in optical lattices. 
In mean-field theory, the Hartree self-energy terms
renormalize $\mu_{o}$ such that $\mu = \mu_o + \mu_U(f) +\mu_V(f)$
where $\mu_U(f)$ and $\mu_V(f)$ are filling-dependent corrections
to $\mu$. We work with the
renormalized $\mu$ so as to properly deal with weak and strong couplings, and
take $\mu_{J_i}(f) = J_if$, where $J_i = U$, $- V$.
The filling $f=N/2M$, with $N$ the number of particles, $M$
the number of lattice sites, and the spin degeneracy factor of 2.
In correlated electron systems, interactions are mainly Coulombic in origin, and $V$ is typically an order of magnitude down from $U$. These are scaled by $t$ for a convenient characterization of weak and strong coupling, and to be broadly applicable. In optical lattices, an {\it effective} Hamiltonian, similar to Eq. (1), could be deduced with $V$ arising from dipolar, or multi-band couplings.

On Fourier transforming and retaining interactions between
particles with equal and opposite momentum, 
%as in BCS theory,
the reduced {\it pairing} Hamiltonian assumes the form:
\begin{equation}
 H_{pair}=\sum_{k}(\epsilon_k-\mu) c^{+}_{k}c_k+\sum_{kk'}V_{kk'}c^{+}_{k'}c^{+}_{-k'}
c_{-k}c_{k}
\end{equation}
where in the tight-binding approximation,
$\epsilon_k=- 2t (\cos k_x+\cos k_y)$; $V_{kk'}=V_0 (\cos(k_x-k'_x)+\cos(k_y-k'_y))$, which
is {\it non-separable}. 
Using the standard BCS variational ansatz,
$|\Phi_{BCS}> = \prod_{\bf k} (u_{\bf k} + v_{\bf k}c_{\bf k}^{{\dag}}
c_{\bf - k}^{{\dag}})|0>$,
we obtain the $T=0$ gap equations
for the gap functions $\Delta_k^{d,s}=\Delta_o (f) (\cos k_x \pm \cos k_y)$
with $d_{x^2-y^2}$(-) and $s^*$(+) symmetries,
 \begin{equation}
\frac{1}{V_o} =
\frac{1}{2 M} \sum_{k}^{\rm{BZ}}\frac{\cos k_x (\cos k_x \pm \cos k_y)}
{E_k^{d,s^*}},
\end{equation}
where
$E_k^{d,s^*} =
((\epsilon_k-\mu)^2+\Delta_{o}^2 (\cos k_x \pm \cos k_y)^2)^{1/2}$.
The Bogoliubov coefficients are given by,
\begin{equation}
|u_{\bf k}|^2; \;\; |v_{\bf k}|^2 = \frac{1}{2}
(1 \pm \frac{\epsilon_k -\mu}{E_k^{d,s^*}}).
\end{equation}
The ratio $v_{\bf k}/u_{\bf k} = - (E_k^{d,s^*} - (\epsilon_k - \mu))/
\Delta_k^{d,s^*}$. 
We readjust $\mu$ for strong
attractions by supplementing the $T=0$ gap equation with the number equation~\cite{leggett}:
\begin{equation}
N= \sum_{k}^{\rm{BZ}} (1-(\frac{\epsilon_{k}-\mu}{E_k^{d,s^*}});
\end{equation}
This determines the self-consistently readjusted $\mu$, which is no longer
fixed at the Fermi level, and makes the gap
equation applicable over the entire range of filling, thereby the BCS and BEC regimes. 
To allow for strong scattering, sums are performed over the entire BZ.
The natural momentum cut-off afforded by the lattice avoids any possible
ultraviolet divergences.
 
Remarkable differences  stem in an essential way from differences in gap symmetry.
The $d_{x^2-y^2}$ gap $\Delta_k^d$ vanishes
along the lines $\pm k_x = \pm k_y$
in the 2D BZ, i.e. at {\sl four} points on the Fermi surface(fs), the location
of which depends upon filling.
The $s^*$ gap $\Delta_k^{s^*}$
coincides with the tight-binding fs at exact 1/2-filling,
and is nodeless otherwise. Here, $\mu \le 0$,
with $\mu = -4t$ at the bottom of the band. Owing to particle-hole
symmetry, it is sufficient to consider $0 \leq  f \leq 1/2$.
The following {\it distinctions} are evident from Eqs. (3-5):
 
(a) For low fillings
($f \rightarrow 0, \mu \rightarrow -4t$), a threshold coupling is required
for d-wave pairing, while in the $s^*$ case,
$\Delta^{s^*} \rightarrow 0$ as $V \rightarrow 0$ due to a
weak singularity at $\mu =-4t$. At 1/2-filling, however, due to a weak singularity at $\mu = 0$
in the d-wave case, $\Delta^d \rightarrow 0$ as $V \rightarrow 0$.
For $s^*$, this singularity is not present, so, as $\Delta^{s^*} \rightarrow 0$, $V/4t \rightarrow \pi^2/8$, i.e.
a minimum coupling is needed for pairing.
In contrast with $\Delta_o^{s^*}(V)$, $\Delta_o^{d}(V)$ changes
slope at $\mu = -4t$, and hence not smooth everywhere (though continuous).
 
(b) For small momenta $k$, the system exhibits the following limiting behavior: 
(i) $\epsilon_k < \mu(= -4t); \;\; |u_k| \rightarrow 1, |v_k| \rightarrow 0$;
this is the strong-coupling BEC limit. Here
the ratio $v_k/u_k \sim \Delta_k/2|\mu| \rightarrow (k_x^2 - k_y^2)/2|\mu|$,
i.e. {\it analytic}.
(ii) $\epsilon_k > \mu(= -4t); \;\; |u_k| \rightarrow 0, |v_k| \rightarrow 1$;
this is the weak-coupling BCS limit. Here
$v_k/u_k \rightarrow 1/(k_x - k_y)$, i.e. {\it non-analytic}.
(iii) $\epsilon_k = \mu(= -4t); \;\; |u_k| \ne 0, |v_k| \ne 0$,
when $E_k \rightarrow 0$.
Then $v_k/u_k  \sim (k_x - k_y)/(k_x + k_y)$, i.e. intermediate between
(i) and (ii).
For d-wave, the quasiparticle excitations
in the BCS limit  (ii) are ``gapless'' for some values
of $k$, while in the  BEC limit (i),
$E_{k} \ne 0$, even for gaps with nodes~\cite{Mohit}.

\begin{figure}
      \centering
      \includegraphics[scale=0.50]{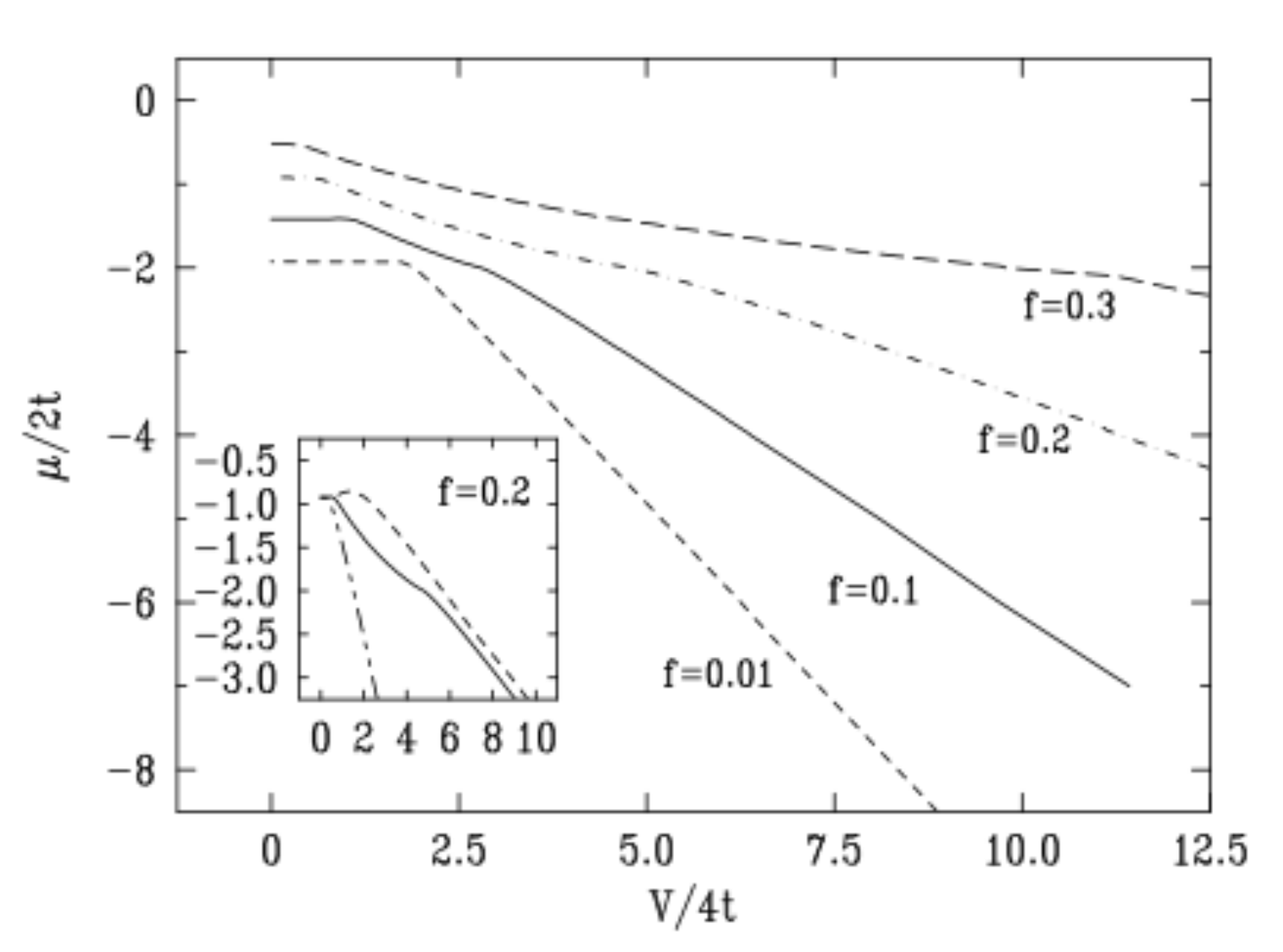}
      \caption{Chemical potential $\mu$ versus coupling $V$  at different
fillings $f$ for $d$-wave pairing.
BEC pairs appear where $\mu(V)$ crosses the $\mu/2t =-2$ line.
The inset shows $\mu(V)$ for $s$- (dash-short dashed line),
$s^*$- (dashed line), and d-wave (solid line) pairing at $f=0.2$.}
\label{fig:Fig.1}
 \end{figure}
 
Self-consistent numerical solutions of Eqs.(3-5) bear out the above features in
detail, and reveal additional features. We scale $\mu$, $V$, $\Delta$
by hopping parameter $t$.
At a given filling $f$, both $\Delta_k^d$ and $\Delta_k^{s^*}$ increase
with increasing $V$. While for d-wave it is
easier to pair electrons at
higher fillings, this is not necessarily the case for
$s^*$-wave for the weaker couplings $V/4t \leq 1.5$ and
small gaps $\Delta^{s^*}/2t \leq 0.5$.

In Fig.1 we show $\mu(V)$ for different fillings f. At a fixed $f$,
in both the $d$- and $s^*$-wave cases, $\mu$ decreases with increasing
coupling $V$, changing less rapidly for progressively larger $f$.
However, as shown in the inset, for $s^*$-wave, $\mu(V)$ exhibits a small ``bump'' for
weaker couplings $V/4t \leq 1.5$; for uniform $s$-wave, the drop in $\mu$ with $V$ is
significantly more rapid.
Crossover to the BEC regime is signaled by
$\mu(V)$ going below the $\mu = - 4t$ line.
%bottom of the band, i.e. crossing the $\mu = - 4t$ line.
As Fig.1 shows, for d-wave, this develops at all fillings for some minimum coupling $V_b/4t$ .
We note that as $f \rightarrow 0$, $V_b/4t \rightarrow 1.8$; at 1/2-filling, this coupling tends to infinitely large values.
For  $V > V_b$, the system is conducive to BEC pairing, and 
for $V < V_b$, the system exhibits BCS-like features.

\begin{figure}[b]
      \centering
      \includegraphics[scale=0.50]{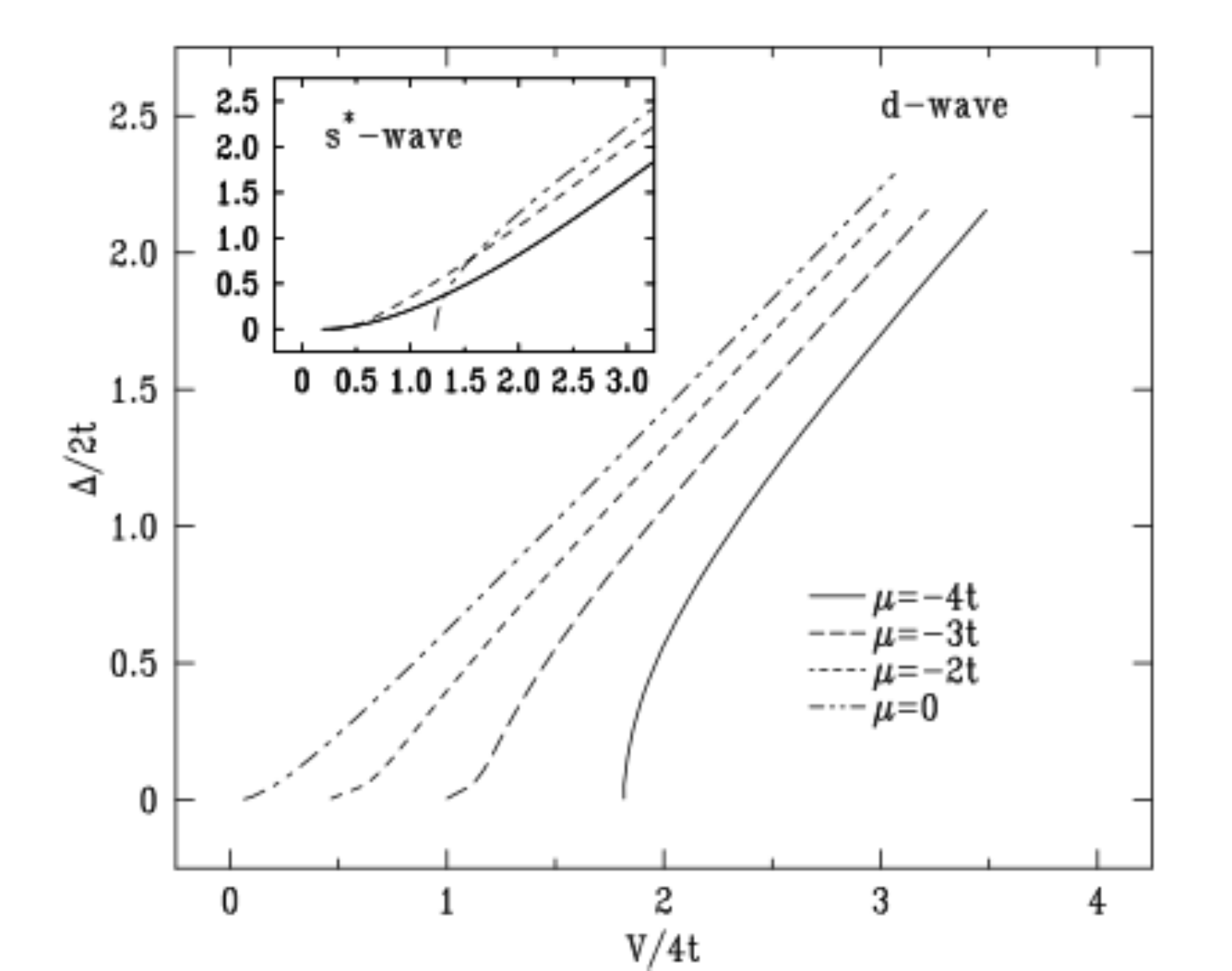}
      \caption{d-wave gap functions $\Delta/2t$ vs. nearest-neighbor coupling
$V/4t$ for different chemical potential $\mu$. Inset: Results for $s^*$ case.
$\mu = - 4t$ demarcates BEC and BCS regimes.}
\label{fig:Fig.2}
 \end{figure}
 
Fig.2 shows the behavior of d-wave gaps as a function of coupling
$V$ for different values of the chemical potential $\mu$.
The $\mu=-4t$ curve represents the locus of $V_b/4t$ for different
fillings (see Fig.1), and demarcates BEC and BCS -pair regimes.
To the left is the $\mu>-4t$ region wherein finite gaps of the BCS or intermediate
BCS-BEC types exist. On a given
constant-$\mu$ curve it may not be possible to have solutions for
any arbitrary filling,
but only those that satisfy Eqs. (3) and (4) self-consistently.
The inset in Fig.2 shows the corresponding $\Delta^{s^*}(V)$ curves
for the $s^*$ case. There are interesting differences with the d-wave results
in that the boundary ($\mu=-4t$) separating BEC/BCS regimes 
is not as clear-cut for weaker couplings $V/4t \leq 1.5$ and
smaller gaps $\Delta/2t \leq 0.5$, however the $\mu<-4t$ region
lies to the right of the $\mu=-4t$ curve, as in the $d$-wave case.

\begin{figure}[t]
     \centering
     \includegraphics[scale=0.40]{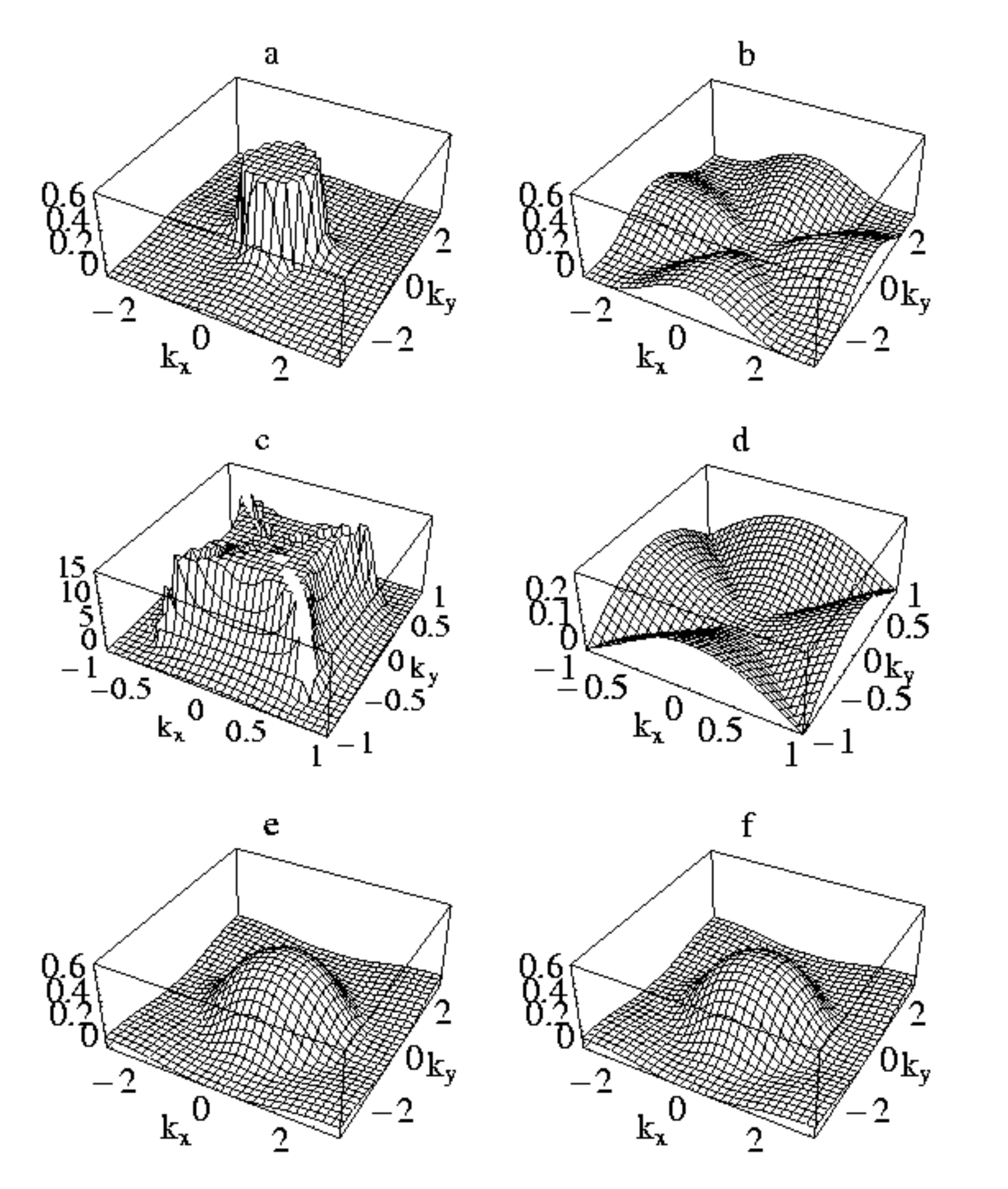}
     \caption{(a), (b): 3D plots of d-wave fermion distribution
functions $v_k^2$ vs. $k_x-k_y$ at filling $f=0.1$, showing
abrupt ``jump'' in $v_k^2$. In
BCS regime (a), $\mu = - 3t$, $\Delta^d = 5.2t$, $ V = 18.7t$, and in
BEC regime (b), $\mu = - 6t$, $\Delta^d = 0.6t$,
$ V = 5.2t$.
(c), (d): 3D plots of d-wave $v_k/u_k$ vs $k_x-k_y$ for the same parameters
as in ((a),(b)) respectively. In BCS regime (c)
it can be seen to be non-analytic; in BEC regime (d)
it is analytic. (e), (f): The same as in (a), (b), but for
$s^*$-wave; the behavior is smooth.}
\label{fig:Fig 3}
     \end{figure}
 
Differences in the gap symmetry manifest in a
striking manner in the momentum distribution function, $v_k^2$, and
the ratio $v_k/u_k$.
For $d$-wave, for a given filling, in the weak-coupling BCS regime
($V < V_b(f)$, $\mu > -4t$), $v_k^2$ exhibits a peak centered
around the zone center (0,0),
that becomes progressively narrow with decreasing filling.
Then at the crossover point
at $V_b(f)$ ($\mu=-4t$), $v_k^2$ {\it abruptly}
goes to zero around (0,0), and shows a drastic redistribution along $(0, \pm \pi)$, 
and $(\pm \pi, 0)$ of BZ. 
The abruptness is manifested in a  ``jump''
in $v_k^2$ as the chemical potential
goes from just above the bottom of the band ($\mu > -4t$) to just
below ($\mu < -4t$), i.e. from BCS to BEC regime. A representative case is shown in Figs. 3a,3b. In marked contrast, for $s^*$-wave, (Figs. 3e, 3f), the zone center peak
in $v_k^2$ decreases {\it smoothly} as one goes from the BCS regime
to the BEC regime; only a slight redistribution occurs at $(\pm \pi, \pm \pi)$.
This behavior is replicated at all fillings $f < 1/2$.
As noted above in the small-$k$ limiting cases, our numerical
calculations show (Fig 3c, 3d) that for d-wave,
in the weak-coupling BCS regime, $v_k/u_k$ is {\it non-analytic}
at $\pm k_x = \pm k_y$; in the strong-coupling BEC regime,
$v_k/u_k$ is {\it analytic},
vanishing along the zone diagonals and peaking about
$(\pm \pi,0)$, $(0, \pm \pi)$.
In the $s^*$ case (not shown), $v_k/u_k$ is analytic in both regimes.
Thus we expect states with  $d$- or $s^*$ pairing gap symmetry to exhibit contrasting behavior at the BCS-BEC crossover, i.e. the unitarity limit.

%Similar behavior in $n_k$ has also been reported in other work~\cite{melo,VQ,derhertog99}.

The Fourier transform of $v_k^2(k_x,k_y)$, namely, $\rho_v(x,y)$ reflect these differences.  In the {\it d-wave case}, in marked contrast 
with its behavior in the BCS regime,
$\rho_v(x,y)$ is {\it oscillatory} in the BEC regime, and exhibits an inhomogenous
``checkerboard-type'' pattern; see Fig 4(a,b).
For the parameters of Fig. 3, the contrast ratio 
of the lowest density to the peak is roughly 50\%, being
most sensitive to the location of $\mu(V)$. The length scale
is of the order of fractions of lattice spacing.
$\rho_v(x,y)$ is fairly uniform in the {\it $s^*$ case} in both regimes.

%\begin{figure}
\begin{figure}[b]
 % \centering
{\scalebox{0.22}{\includegraphics[clip,angle=0]{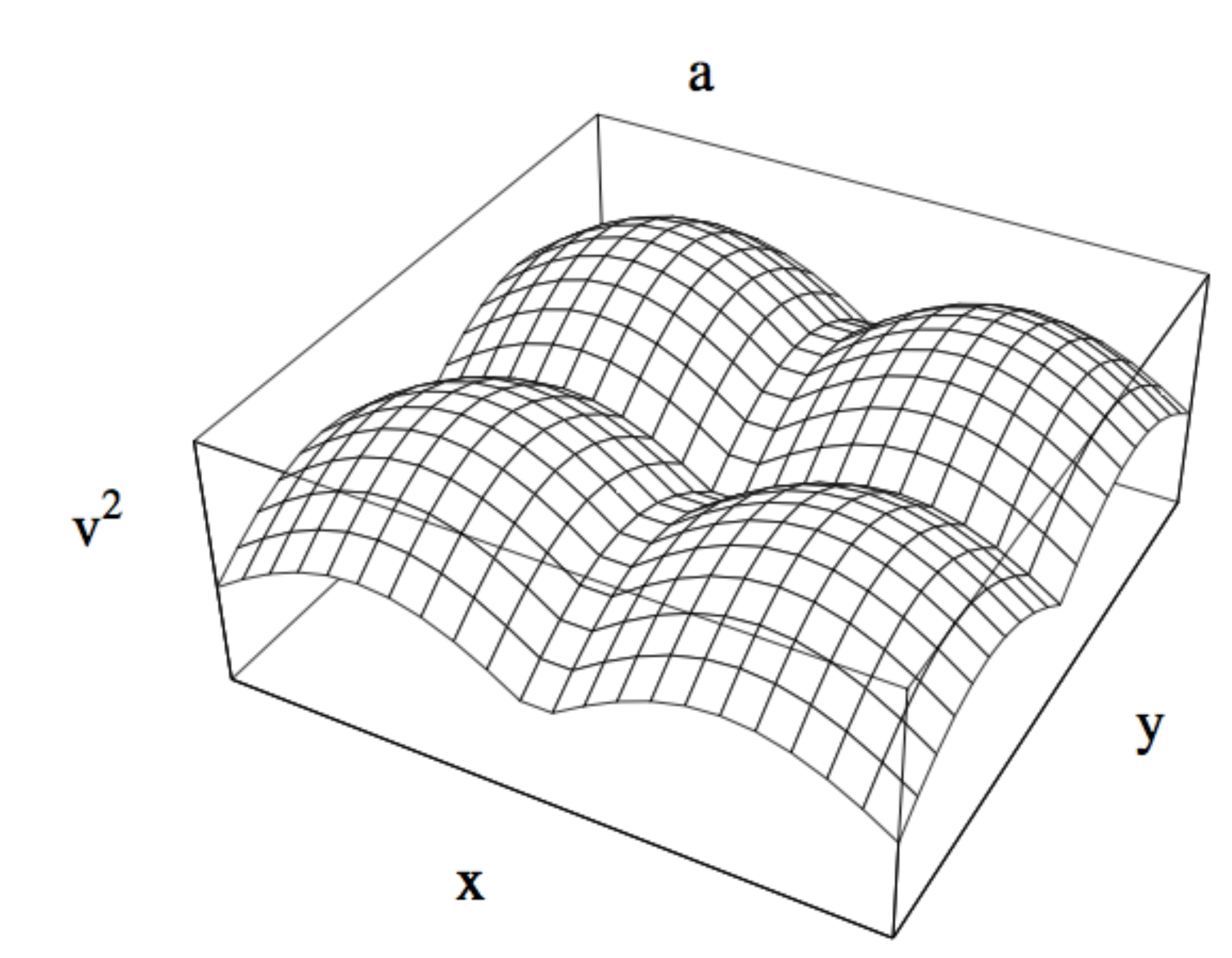}}}
{\scalebox{0.22}{\includegraphics[clip,angle=0]{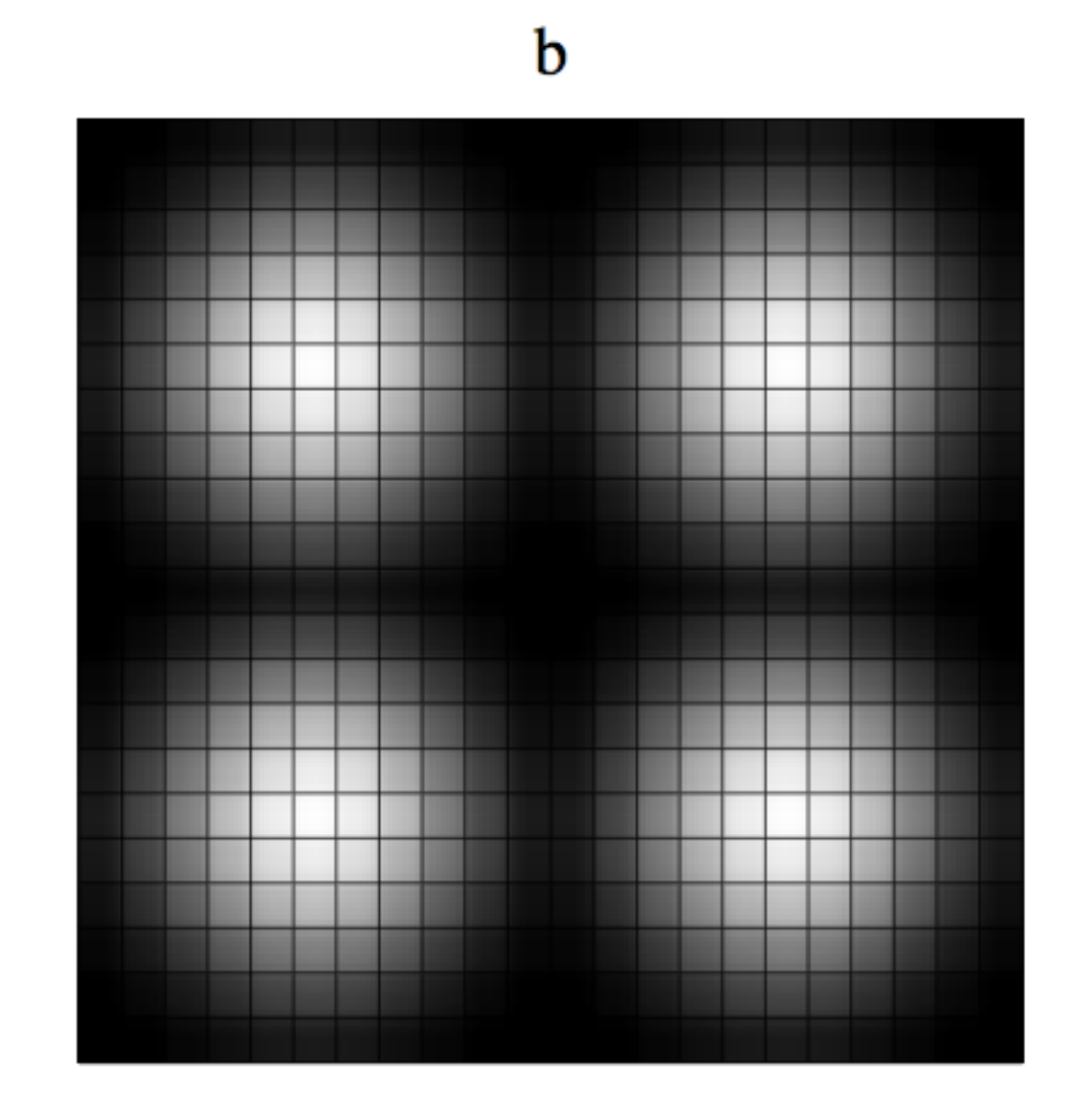}}}
     \caption{(a) Fourier transform $\rho_v(x,y)$ of a typical d-wave
fermion distribution function, $v_k^2$. Here, filling $f=0.01$,
$\mu = - 4.2t$ (strong-coupling regime), gap $\Delta = .76t$. (b) Projection
of (a) to show contrast ratio of $\rho_v(x,y)$.}\label{fig:Fig 4}
\end{figure}
  %   \end{figure}
  
For correlated electrons, existent experimental techniques can reveal the proposed signatures to
distinguish between states with $d$ or extended-s ($s^*$) symmetry;
e.g. $v_k^2$, its Fourier transform, 
%related to quasiparticle momentum distribution, and its Fourier transform, 
and quasiparticle energy could be deduced from ARPES~\cite{Shen}, pair symmetries from quasiparticle tunneling or scanning tunneling microscopy (STM)~\cite{STM}. In cold fermion systems, $v_k^2$, and its Fourier transform, could be determined from time-of-flight~\cite{tof} measurements, in which atoms are released from the lattice, and imaged at a later time.
Combined with rf pairing gap spectroscopy~\cite{rf-pair,lambda} (analog of tunneling), this could decipher the pairing symmetry. Momentum resolved rf spectroscopy~\cite{rf,feld},
could provide information on the quasiparticle energy $E_k = (\Delta_k^2 + \epsilon_k^2)^{1/2}$, thereby shedding light on ($u_k,v_k$) and density of states.
%Highly sensitive STM may be able to pick up such distinctions\cite{STM}.
% orcoherence factors,   $u_kv_k +u_k'v_k'$.
%Angle-dependent or transverse ultrasound attenuation\cite{coleman},
%or quasiparticle tunneling at low fillings
%are possible experiments.

Here we have considered a homogeneous Fermi system, so as to appeal broadly to both
condensed matter and cold atom physics.  While we expect key aspects of our results to hold for
cold fermion systems in the presence of atomic traps, we comment on possible effects of the trap. But, first, we note that we consider  a pairing Hamiltonian based 
on the extended Hubbard model, and not the simple Hubbard model, on which most discussions in cold atom literature are based. We are also away from 1/2 filling, and at relatively strong coupling, so that possible effects of spin density wave (SDW) and charge density wave (CDW) instabilities are expected to be suppressed. We have checked~\cite{VQ} that addition of next-near-neighbor hopping tends to  stabilize the paired state, as well as lower the minimum
near-neighbor interaction necessary for a bound-state.

Away from 1/2 filling (lower fillings) for a 2D lattice in presence of trap (within local density approximation (LDA)), the multitude of quantum phases obtained~\cite{Drummond,Kaponen,Heisel} near 1/2 filling may not exist~\cite{Drummond, Kaponen}, leaving a metallic state from the trap center to the edge. Thus, for this range of fillings, our results are not expected to be subject to competing signatures from other possible phases in rf spectra. Closer to 1/2 filling, within LDA, there may be other phases, e.g. pairing in different shells behaving as infinite system in each shell~\cite{Heisel}.  Then, tomographic (spatially resolved)  
rf spectroscopy~\cite{tomog}, using {\it in situ} phase contrast imaging technique, would be able to probe each shell region, within each of which our results should hold.

%As standard for homogenous systems, we have absorbed the Hartree term in the chemical potential. 
Recent work~\cite{Giam} have pointed out possible effects of trap inhomogeneity on the Hartree term, and consequently on rf spectra. In this work, Hartree effect on self-energy is obtained at first-order, so the role of higher-order terms towards a convergent result is not clear at this point. It may also be interesting to examine Hartree effect for the extended Hubbard model on which our calculation is based; this is outside the scope of current work.

Our calculation is at T= 0, and any Berezinski-Kosterlitz-Thouless transition~\cite{botelho}  would only be revealed in a finite-T calculation. Also, like others, we take our the system  
to be not strictly 2D, but quasi-2D, with the assumption that a weak link along the 3rd direction stabilizes phase transitions like superfluidity. Our T=0 consideration does not lend itself to calculations of critical 
temperature, $T_c$. However,
based on other work~\cite{Bickers,hofstetter02}, for d-wave pairing we estimate that in electron systems, such as the cuprates, $T_c/T_F \sim 0.015- 0.03$, giving a $T_c \sim 15 K - 30 K$ for a $T_F \approx 10^3 K$;
in cold fermions, $T_c/T_F \sim 0.01$, giving a $T_c \sim  30 nK$ for a $T_F \approx 3 \mu K$. Thus for
d-wave $T_c$ measurements in cold fermions, realizing temperatures below the currently  attainable $ T/T_F \simeq 0.05$ is needed.  However, these are lower bound estimates, and we expect our proposed signatures to persist to higher temperatures. 
%Recent measurements\cite{lambda} of lambda transition at $T_c/T_F = 0.16$, together with 
Recent suggestions~\cite{cooling} of novel cooling methods are encouraging.

Our calculations, in the spirit of BCS and 
BEC-BCS crossover theories, consider  $d$- and extended-s ($s^*$) wave pairing symmetries, independent of pairing mechanisms. Though mean-field in nature, 
we expect the calculated signatures of unconventional pairing symmetries to hold in calculations beyond mean fields. A recent Monte Carlo work~\cite{trivedi} on fermions on 2D optical lattice, though at 1/2 filling, goes beyond mean field and contain substantial discussions regarding possible phases.

%We have not explored here the issues of collective modes 
%or phase separation\cite{phasesep}.
%It may be interesting to
%extend this work to, for example, finite-T, or to explore whether
%the inhomogenous density that we find bear relationship to
%the range/strength  of the interaction, or to possible phase separation.
 
We thank E. Abrahams, S. Davis, and H. Neuberger for discussions, and acknowledge the support of Aspen Center for Physics, where part of the work was carried out. 
The work was supported in part by the Institute for Complex Adaptive Matter (ICAM).

\end{document}